\documentclass[pre,aps,twocolumn,amsfonts,amssymb,amsmath,eqsecnum,floatfix]{revtex4}
\usepackage{bm}
\usepackage{graphicx}
\usepackage[dvips]{color}
\usepackage{amsfonts}
\usepackage{amsmath}
\usepackage{amssymb}
\usepackage{amsbsy}
\usepackage{epsfig}
\usepackage{ulem}
\newcommand{\symmtraceless}[1]{\vbox{\ialign{##\crcr
       \vrule height0.4pt depth2pt
       \hrulefill
       \vrule height0.4pt depth2pt
       \crcr\noalign{\kern1pt\nointerlineskip}
       $\hfil\displaystyle{#1}\hfil$\crcr}}}
\newcommand{\stl}[1]{\symmtraceless{#1}}
\newcommand{\lambdaK}{\lambda_{\mathrm{K}}}

\usepackage{bm}

\begin{document}
\title{Shear-stress controlled dynamics of nematic complex fluids}

\author{Sabine H.~L.~Klapp and Siegfried Hess}
\affiliation{
Institut f\"ur Theoretische Physik, Sekr.~EW 7--1,
Technische Universit\"at Berlin, Hardenbergstrasse 36,
             D-10623 Berlin, Germany}
\date{\today}

\begin{abstract}
Based on a mesoscopic theory we investigate the non-equilibrium dynamics of a sheared nematic liquid, with the control parameter being the shear stress
$\sigma_{\mathrm{xy}}$  (rather than the usual shear rate, $\dot\gamma$). To this end we supplement the equations of motion for the orientational order parameters by an equation 
for $\dot\gamma$, which then becomes time-dependent.  Shearing the system from an isotropic state, the stress-controlled flow properties turn out to be essentially identical to those
at fixed $\dot\gamma$. Pronounced differences occur when the equilibrium state is nematic. Here, shearing at controlled $\dot\gamma$ 
yields several non-equilibrium transitions between different dynamic states, including chaotic regimes. The corresponding stress-controlled system has only one transition
from a regular periodic into a stationary (shear-aligned) state. The position of this transition
in the $\sigma_{\mathrm{xy}}$-$\dot\gamma$ plane turns out to be tunable by the delay time entering our control scheme for $\sigma_{\mathrm{xy}}$.
Moreover, a sudden change of the control method can {\it stabilize} the chaotic states
appearing at fixed $\dot\gamma$.
\end{abstract}
\pacs{???}
\maketitle
\date{\today}

\section{Introduction}
\label{intro}
Shearing a fluid of orientable, e.g. rod-like particles is a fundamental example of a soft-matter system driven far away from equilibrium. The
complex dynamics of such systems, examples being worm-like or cylindrical micelles \cite{Schmitt94,Berret02}, (polymeric) liquid crystals, blood cells, or colloidal rod (e.g. tobacco virus) suspensions, has become a focus of many experimental investigations \cite{Larsonbook}, computer simulations \cite{Tao05,Tao06,GERM05,Ripoll08}, and  theoretical studies
over the last years (for a recent review, see \cite{FIEL07}). Within the isotropic high-temperature (or low-concentration) state, application of external shear 
typically yields either continuous flow-alignment or a discontinuous paranematic-nematic transition \cite{Lettinga05}, depending on how close the underlying equilibrium transition
is \cite{Olmsted92,Hess76a}. Corresponding rheological properties such as the ''constitutive curve" formed by the shear stress $\sigma_{\mathrm{xy}}$ as function of the shear rate $\dot\gamma$
are already non-linear, implying a non-Newtonian behavior of the viscosity. More spectacularly, shearing inside the nematic (i.e., orientationally ordered, yet translationally disordered) phase can induce various bifurcations, that is, non-equilibrium transitions between different dynamic ''states'' characterized by a specific periodic or non-periodic motion of the nematic director
\cite{Rien02,Rien02a,Grosso03,Forest04,Ripoll08}. Examples are the tumbling or wagging behavior observed both by theory \cite{Rien02} and in experiments of sheared tobacco viruses \cite{Lettinga05}.
Further issues receiving much attention
are the appearance of rheochaos (chaotic stress-strain curves and/or orientational dynamics) \cite{Sood00,Sood01,Rien02,Rien02a,Grosso03,Fielding04,ARA05,GOD08}, 
and, sometimes combined with that, spontaneous spatial-symmetry breaking associated to shear-banding \cite{FIEL07,LER08}. In the latter situation, the system separates into domains characterized
by different local shear rates.

Many of these fascinating phenomena can be described by mesoscopic theories focussing on the motion of some suitable, coarse-grained dynamic variables. An example
is the approach developed by Hess \cite{Hess75,Hess76b} and Doi \cite{Doi80,Doi81},
which is based on an equation of motion for the tensorial orientational order parameter, that is, the alignment tensor ${\bf a}(t)$. A constitutive relation then 
yields the link from ${\bf a}(t)$ to the stress tensor ${\bm\sigma}$ and, thus, the rheology of the system. The theory is highly non-linear due 
to the use of a fourth-order Landau-de Gennes free energy functional describing the relaxation of ${\bf a}$ towards equilibrium. 
As a consequence of this non-linearity, the (five-dimensional) dynamical system involving the independent components of ${\bf a}$ generates complex orientational dynamics and associated rheological properties,
even if the analysis is restricted to simple (Couette) shear geometries and to spatially {\it homogeneous} systems (for extensions of the Hess-Doi approach towards inhomogeneous systems
see, e.g., \cite{Das05,Heidenreich09,Hess81,Borg95}). Another class of mesoscopic theories focusses directly on the shear stress $\sigma_{\mathrm{xy}}$ as a dynamic variable
\cite{FIEL07,ARA05,Cates02,Aradian06,GOD08}, an example being the non-local Johnson-Segalman model \cite{John77,Fielding04}. These models are capable of describing, on a quite general level,
complex rheological behavior such as shear-banding \cite{FIEL07}, a drawback being that they reflect the (orientational and/or translational) dynamics within the underlying liquid only indirectly.

In the present study we are interested in both, rheological properties and orientational motion, of a complex fluid composed of rod-like particles. So far, most theoretical studies investigating the orientational dynamics take the shear rate $\dot\gamma$ as an external driving parameter. Experiments, on the other hand,
are often conducted at controlled shear stress (see, e.g., \cite{Hu98,Volkova99,Herle05}); moreover, there are various devices where either stress or strain can be controlled. Thus, it is an interesting question
to which extent these macroscopic conditions influence the observed dynamics. 

In the present study we explore the implications of controlled stress (as compared to controlled shear rate) via an extension of the mesoscopic equations for ${\bf a}(t)$ \cite{Hess75,Hess76b}.  
In order to fix the (time-averaged) shear stress at an externally imposed value, we 
supplement the five equations of motion for the nematic order parameters by a further differential equation for $\dot\gamma$.
The approach of $\sigma_{\mathrm{xy}}$ towards its target value is governed by a control time, $\tau_{\mathrm{g}}$. This control scheme 
is inspired by the method used in experimental devices: 
the shear rate is adjusted such that the desired value of the shear stress is approached. Some preliminary results for stress-controlled systems have been published in Ref.~\cite{Hess95}.
In the present paper we undertake a more systematic study focussing on spatially homogeneous states with only one (constant) non-vanishing component
of the shear rate (and stress-) tensor. This restriction excludes, e.~g., the appearance of shear bands
\cite{FIEL07,LER08}, shear-induced layering \cite{Rendon07}, and other effects \cite{Park09} such as the emergence of secondary flow.
However, we consider an analysis of the homogeneous system as compulsory before the full hydrodynamics with spatially inhomogeneous shear rates is performed. Nonetheless, the present control approach could, in principle, be applied to inhomogeneous systems (and cross-couplings between deformation rate and stresses)
without further ado. 
Indeed, the mesoscopic theory on which our study is based, is fully capable of treating inhomogeneous effects, and we have already performed
investigations in this direction (yet without stress control) in Refs.~\cite{Heidenreich09,Heidenreich07,Heidenreich08,Forest08}. 
A similar strategy was taken by other researchers (see, e.g. \cite{Tsuji98}).

Numerical results for both, orientational dynamics and rheological properties (i.e., stress-strain curves and resulting viscosities)
are presented for stress-controlled systems at a fixed value of the ''tumbling'' parameter \cite{Larsonbook,Hess76b} (that is, 
the coupling strength between flow and orientational motion), which seems particularly promising based on earlier studies under fixed-strain conditions \cite{Rien02}.
The most interesting results are found at temperatures below the isotropic-to-nematic (I-N) phase transition, where the control method turns out to be crucial for the dynamics of the system. 
This concerns both the type of bifurcations and their characteristics. Moreover, we show how our control scheme can be used to change {\it chaotic} states
seen in systems without stress control into states with regular dynamics, or into stationary states. 
This puts our study in the more general context of control in non-linear dynamic systems, where the stabilization of steady or periodic fix points by
feedback control of suitable dynamic variables is a major issue \cite{Schoellbuch}.

The paper is organized as follows. In the subsequent sections~\ref{general}-\ref{explicit_equations} 
we summarize the main ingredients of the mesoscopic theory for rod-like particles under shear flow. Our control
method is introduced in Sec.~\ref{shear_control}. Section~\ref{results} describes our numerical results, with an emphasis on sheared nematic systems. Finally, in Sec.~\ref{conclude} we 
summarize our main findings
and suggest some future directions of research.
\section{Theory}
\label{theory}
\subsection{Order parameter and dynamic equations}
\label{general}
We employ a mesoscopic description 
of the system, where the relevant dynamic variable is the orientational order parameter averaged over some volume in space. In a sheared liquid crystal, this order parameter corresponds to the time-dependent, 2nd-rank alignment tensor ${\bf a}=\sqrt{15/2}\langle \stl{{\bf u}{\bf u}}\rangle$, where 
${\bf u}$ describes the molecular axis and $\stl{\ldots}$ indicates the symmetric traceless part of a tensor.
In the isotropic equilibrium state, all components of ${\bf a}$ are zero. For {\it uniaxial} nematic orientational order, 
${\bf a}=a\sqrt{3/2}\langle \stl{{\bf n}{\bf n}}\rangle$ with $a=\sqrt{5}S$ being a scalar proportional to the Maier-Saupe order parameter $S$, and ${\bf n}$
being the corresponding director. In a sheared system, however, the assumption of uniaxial order (which leads to the so-called Ericksen-Leslie theory of orientational dynamics
\cite{Ericksen61,Leslie68})
remains appropriate only for small shear rates and temperatures deep within the nematic phase. For intermediate and large shear rates and, in particular, in the vicinity of the 
I-N phase transition, biaxial ordering occurs, and a complete description of the non-equilibrium dynamics  
requires the full tensor, ${\bf a}(t)$. 

Taking ${\bf a}$ as the relevant variable, the system's equilibrium behavior (in the absence of shear) is governed by the Landau-de Gennes free energy \cite{deGennes}
\begin{equation}
\label{free_energy}
 \Phi=\frac{1}{2}A(T){\bf a}:{\bf a}-\frac{\sqrt{6}}{3}B\left({\bf a}\cdot{\bf a}\right):{\bf a}+\frac{1}{4}C\left({\bf a}:{\bf a}\right)^2,
 \end{equation}
where the notation ":" stands for the trace over the product of the two symmetric tensors, and "$\cdot$" indicates conventional matrix multiplication. In Eq.~(\ref{free_energy}), $T$ is the 
temperature which is the driving parameter for an I-N transition in a thermotropic liquid crystal (for lyotropic systems, one would
replace $T$ by the concentration of the particles).

Switching on an external shear
flow characterized by a velocity field ${\bf v}$, the alignment tensor becomes a time-dependent quantity. Here we employ the approach first 
derived from concepts of irreversible thermodynamics \cite{Hess75} and later from a generalized 
Fokker-Planck equation for the orientational distribution function \cite{Hess76b,Doi80,Doi81} (for similar though not identical approaches, see, e.g., \cite{Ericksen61,Stark03}).
For homogeneous systems the equation of motion reads
\begin{equation}
\label{eq:Hess}
\frac{\partial {\bf a}}{\partial t}
-2\stl{{\bm \omega}\times{\bf a}}
- 2\kappa\,\stl{{\bm \Gamma}\cdot{\bf a}}
+ \tau_{\mathrm{a}}^{-1}\,{\bf \Phi}({\bf a})
  = -\sqrt{2}\,\frac{\tau_{\mathrm{ap}}}{\tau_{\mathrm{a}}}\,{\bm \Gamma},
\end{equation}
where ${\bm \omega}=(1/2)\nabla\times{\bf v}$ is the vorticity and
${\bm \Gamma}=\stl{\nabla {\bf v}}$ is the strain rate tensor. Specializing to the case of a plane Couette flow
characterized by a flow in $x$-direction, ${\bf v}=\dot\gamma y {\bf e}_{\mathrm{x}}$, the strain rate tensor simplifies
to ${\bm \Gamma}=\dot\gamma\stl{{\bf e}_{\mathrm{x}}{\bf e}_{\mathrm{y}}}$ and 
${\bm \omega}=-(1/2)\dot\gamma {\bf e}_{\mathrm{z}}$. In these formulae, ${\bf e}_{\alpha}$ ($\alpha=x,y,z$) denote unit vectors
along the coordinate axes. 
Further,  the parameters $\tau_{\mathrm{ap}}$ and $\tau_{\mathrm{a}}$ are relaxation time coefficients, and $\kappa$ is a dimensionless coefficient.
Finally, the (tensorial) quantity ${\bf \Phi}({\bf a})$ appearing in Eq.~(\ref{eq:Hess}) corresponds to the derivative
of the free energy~(\ref{free_energy}) with respect to the (non-conserved) order parameter, i.e., ${\bf \Phi}=\partial \Phi/\partial {\bf a}$. In the absence of shear (i.e., for ${\bf v}={\bm \omega}={\bm \Gamma}=0$)
this terms governs the relaxation of ${\bf a}(t)$ towards its equilibrium value determined by  ${\bf \Phi}=0$. We note that the Ericksen-Leslie theory of orientational dynamics
\cite{Ericksen61,Leslie68}
follows from the present approach, when the alignment tensor is uniaxial and the 
effect of the shear flow on the magnitude of the order parameter can be disregarded. Furthermore, as noted already in the Introduction,
the present approach can be generalized and has been already applied to inhomogeneous alignment states 
\cite{Heidenreich09,Heidenreich07,Heidenreich08,Forest08}, central steps being the inclusion of gradient elasticity terms in the free
energy  (see also \cite{Tsuji98}) and an appropriate description of feedback effects of the orientational motion on the flow profile 
\cite{Heidenreich09,Heidenreich07,Heidenreich08,Forest08}. Not surprisingly, however, these generalizations of the theory imply a substantial
increase of the numerical effort, which is the reason why we stick to homogeneous states in the present work.

We next consider the rheological properties of the sheared liquid crystal, which are characterized by its 
pressure tensor, ${\bf p}$ \cite{Hess95}. The latter consists of an isotropic part involving the hydrostatic pressure $p$,
an antisymmetric part, and an symmetric traceless part
$\stl{{\bf p}}$ which we refer to as friction pressure tensor \cite{Hess75}. The negative of this latter contribution
is the so-called shear stress tensor, ${\bm \sigma}=-\stl{{\bf p}}$.
This quantity is of special interest since it is directly linked with the
orientational dynamics in the non-equilibrium system. Specifically, one has
\begin{equation} 
\label{eq:HessEqPressure}
\stl{\bf p} = -2\eta_{\mathrm{iso}}
  {\bm \Gamma} + \stl{{\bf p}_{\mathrm{al}}},
\end{equation}
where $\eta_{\mathrm{iso}}$ is the so-called second Newtonian viscosity, and
\cite{Borg95}
\begin{equation}
\label{p_al}
\stl{{\bf p}_{\mathrm{al}}}=\frac{\rho k_{\mathrm{B}}T}{m}
\left(\sqrt{2}\frac{\tau_{\mathrm{ap}}}{\tau_{\mathrm{a}}}
{\bf \Phi}({\bf a})-2\kappa\stl{{\bf a}\cdot{\bf \Phi}({\bf a})}\right).
\end{equation}
In Eq.~(\ref{p_al}), $\rho$ is the density of the rod-like particles, $m$ is their mass, and $k_{\mathrm{B}}$ is Boltzmann's constant.
In the absence of shear, one has ${\bf \Phi}({\bf a})=0$ and consequently $\stl{{\bf p}_{\mathrm{al}}}=0$. The friction pressure tensor then reduces to the
first term in Eq.~(\ref{eq:HessEqPressure}), which is present also in (sheared) systems of spherical particles.
A concept to {\it control} the shear stress externally is formulated in Sec.~\ref{shear_control}.
\subsection{Explicit equations of motion}
\label{explicit_equations}
In the practical analysis it is convenient to use scaled variables. Details can be found, e.g., in Refs. \cite{Hess75,Hess81,Borg95} and \cite{Grandner07}. 
The equilibrium behavior of the system is determined by the effective temperature, 
$\theta=9AC/2B^2$. It follows that the (first-order) I-N transition occurs at $\theta=1$ (which corresponds to the I-N coexistence point), and that
the nematic phase (isotropic) is (meta-)stable for temperatures $\theta<9/8$ ($\theta>0$).
For the description of the non-equilibrium system,
times and shear rates are made dimensionless with
a convenient reference time. The latter is chosen as the relaxation time of the alignment at the isotropic-nematic coexistence, i.e.,
$\tau_{\mathrm{ref}} = \tau_{\mathrm{a}} (9C/2B^2)$. Further, the strength of the coupling between the flow and the alignment tensor is characterized by the parameter
$\lambda_{\mathrm{K}}= -(2/3)\sqrt{3} \tau_{\mathrm{ap}}/\tau_{\mathrm{a}}$.  Microscopically, $\lambda_{\mathrm{K}}$ is related to the 
{\it shape} of the particles, which can be characterized by the axis ratio 
$q$ (for ellipsoidal particles). Specifically, one has \cite{Hess76b} 
$\lambda_{\mathrm{K}}=2/(\sqrt{5}a_{\mathrm{K}})(q^2-1)/(q^2+1)$
 where $a_{\mathrm{K}}=2B/3C$. It follows that $\lambda_{\mathrm{K}}=0$ 
for spherical particles ($q=1$), whereas 
$\lambda_{\mathrm{K}}>0$ for elongated particles ($q>1$), which is the case considered here. Finally, the scaled analogue of the
shear rate $\dot\gamma$
is denoted as $\Gamma=\tau_{\mathrm{ref}}\dot\gamma$ (with $\tau_{\mathrm{ref}}$ defined above).

Besides scaling, the other important step towards practical solution of the problem 
is to rewrite the tensorial equation~(\ref{eq:Hess}) for the dynamics of ${\bf a}$ into a set 
of scalar equations for its five independent components, $a_0,a_1,\ldots,a_4$. 
This is achieved via an expansion of ${\bf a}$ in a standard tensorial basis \cite{Rien02a,Kaiser92}. The resulting components are linked by simple expressions to their cartesian counterparts 
$a_{\alpha\beta}$ (with $\alpha,\beta=x,y,z$), that is, $a_0\propto -(a_{\mathrm{xx}}+a_{\mathrm{yy}})$, $a_1\propto a_{\mathrm{xx}}- a_{\mathrm{yy}}$, 
$a_2\propto a_{\mathrm{xy}}$, $a_3\propto a_{\mathrm{xz}}$, and $a_4\propto a_{\mathrm{yz}}$ \cite{Rien02a,Kaiser92}. Thus, $a_0$ (with negative values), $a_1$, and $a_2$ describe ordering within the shear ($x$-$y$) 
plane, whereas $a_3$ and $a_4$ describe out-of-plane ordering. 
The ordinary (first-order) differential equations for the new components of ${\bf a}$ are given by
\begin{eqnarray}
\label{eq:T5Dyn}
  \dot a_0 & = & -  \phi_0  - \frac 1 3\sqrt 3\, \kappa \, \Gamma \, a_2 \,  \nonumber \\
  \dot a_1 & = & -  \phi_1 + \Gamma \, a_2 \nonumber\\
  \dot a_2 & = & -  \phi_2 -   \Gamma \,  a_1 
  + \frac{\sqrt{3}}{2}\lambdaK \, \Gamma  - \frac 1 3\sqrt 3\, \kappa \, \Gamma \, a_0 \nonumber\\
  \dot a_3 & = & -  \phi_3 + \frac 1 2 \Gamma \, (\kappa + 1) \, a_4\nonumber \\
  \dot a_4 & = & -  \phi_4 + \frac 1 2 \Gamma \, (\kappa - 1) \, a_3,
\end{eqnarray}
where the quantities $\phi_0,\ldots,\phi_4$, which correspond to the (dimensionless) components of the free energy derivative
${\bf \Phi}({\bf a})$, are given as
\begin{eqnarray}
\label{phicomponents}
 \phi_0 & = & (\theta - 3 a_0 + 2 a^2) a_0 + 3 (a_1^2 + a_2^2) - \frac 3 2 (a_3^2+a_4^2)  \nonumber \\
 \phi_1 & = & (\theta + 6 a_0 + 2 a^2) a_1 - \frac 3 2\sqrt 3 (a_3^2 - a_4^2)  \nonumber \\
 \phi_2 & = & (\theta + 6 a_0 + 2 a^2) a_2 - 3 \sqrt 3\, a_3 a_4 \nonumber\\
 \phi_3 & = & (\theta - 3 a_0 + 2 a^2) a_3 - 3\sqrt 3 (a_1 a_3 + a_2 a_4)  \nonumber\\
 \phi_4 & = & (\theta - 3 a_0 + 2 a^2) a_4 - 3\sqrt 3 (a_2 a_3 - a_1 a_4).
\end{eqnarray}
In Eqs.~(\ref{phicomponents}) we have introduced the abbreviation $a^2 \equiv\sum_{k=0}^4 a_k^2$.

As outlined before in Sec.~\ref{general} the rheological properties of the system are described by the 
friction tensor or, equivalently, the
stress tensor, ${\bm \sigma}$. Introducing the reference shear modulus $G_{\mathrm{al}}$, we find from Eqs.~(\ref{eq:HessEqPressure}) and (\ref{p_al}) \cite{Hess95}
\begin{equation} 
\label{eq:HessEqPressurezw}
 G_{\mathrm{al}} \, {\bm \sigma} = - \stl{{\bf p}} = 2 \, \eta_{\mathrm{iso}}
   {\bm \Gamma}- \stl{{\bf p}_{\mathrm{al}}} = 
2\eta_{\mathrm{iso}} {\bm \Gamma} +  
\sqrt{2}  G_{\mathrm{al}} \,{\bm \Sigma}^{\mathrm{al}},
\end{equation}  
where
\begin{equation}
{\bm \Sigma}^{\mathrm{al}}\equiv \frac{2}{\sqrt{3}} \lambda_{\mathrm{K}}^{-1} 
\left(\bm \Phi +
        \frac{2\kappa}{3 \lambda_{\mathrm{K}}} \sqrt{6} \,
        \stl{{\bf a}\cdot{\bm \Phi}}\right).
\label{Sigma}
\end{equation}
Here we are particularly interested in the {\it shear stress}, that is, the $x$-$y$ component of ${\bm\sigma}$.
Expanding ${\bm \sigma}$ and ${\bm \Sigma}^{\mathrm{al}}$ into basis tensors one obtains from Eq.~(\ref{eq:HessEqPressurezw}) \cite{Hess95}
\begin{equation}
\label{shearstress}
\sigma_{\mathrm{xy}} = \eta_{\mathrm{iso}} \Gamma + \Sigma_2,
\end{equation}
where
\begin{eqnarray}
 \Sigma_{2}&=&\frac{2}{\sqrt{3}\lambda_{\mathrm{K}}}\left[ \phi_2
 - \tilde{\kappa} \left(
a_2\phi_0+a_0\phi_2\right)\right]\nonumber\\
& & - \tilde{\kappa} (a_4\phi_3+a_3\phi_4) 
\label{Ncomponents}
\end{eqnarray} 
with $\tilde{\kappa} = 2 \kappa /(3 \lambdaK)$.
\subsection{Controlling the shear stress}
\label{shear_control}
Previous applications of the dynamical equations (\ref{eq:T5Dyn}) have focussed on the system's orientational dynamics and rheological behavior 
at constant shear rate, $\Gamma$. The main goal of the present work is to explore the ''reverse'' situation where, instead of $\Gamma$, the shear stress
is held fixed at an (externally imposed) value $\sigma_{\mathrm{xy}}^{\mathrm{imp}}$. 
Indeed, rheological experiments of complex fluids often involve the shear stress as an external control parameter rather than the shear rate  \cite{Hu98,Volkova99,Herle05}.

Within our theory,  the {\it instantaneous} value of $\sigma_{\mathrm{xy}}$ is determined by $\Gamma$ and the instantaneous values for the
alignment components $a_k(t)$ [see Eqs.~(\ref{shearstress}) and (\ref{Ncomponents})]. In order to {\it control} the shear stress we follow the experimental strategy and 
adjust the shear rate, that is, $\Gamma$ becomes a time-dependent quantity. This adjustment is achieved such that we 
supplement the five differential equations for $a_k(t)$ by an additional equation for $\Gamma(t)$. Specifically,
\begin{equation}
\label{dyngam}
\tau_{\mathrm{g}} \, \frac{d \Gamma}{dt} = - \frac{1}{\eta_{\mathrm{iso}}}\left(\sigma_{\mathrm{xy}}(t) - \sigma^{\mathrm{imp}}_{\mathrm{xy}}\right).
\end{equation}
Here $\sigma_{\mathrm{xy}}$ is the instantaneous shear stress as given by (\ref{shearstress}),  $\eta_{\mathrm{iso}}$ is the (second Newtonian) viscosity 
discussed above, and $\tau_{g}$ is a control (or ''delay")  time determining the speed of shear stress control. 
\subsection{Numerical solution}
\label{numerics}
In order to solve numerically 
the six-dimensional dynamical system consisting of Eqs.~(\ref{eq:T5Dyn}) for $a_k(t)$ ($k=0,\ldots,4$)
and Eq.~(\ref{dyngam}) for the shear rate $\Gamma(t)$, we employ a standard four-step Runge-Kutta algorithm with adaptive step size control. The resulting time steps are in the range
$\Delta t=0.001$ - $0.5$.
The initial values for $a_0,\ldots,a_4$ are set to small, yet unzero values (between $0.01$ and $0.1$)
corresponding to a state with weak orientational order. The
initial  shear rate is set to  $\Gamma(0) = 0.01$,  if not stated otherwise.

In analogy to rheological experiments, which are typically not time-resolved, 
we average the results for $\Gamma$ (at controlled stress conditions) or $\sigma_{\mathrm{xy}}$ (controlled shear rate), respectively,
over an appropriate number of time steps. To this end we disregard the initial transient behavior, which may strongly depend on the start conditions.
The overall number of time steps required to obtain reliable averages strongly depends on the system parameters considered. In the present work we consider a variety
of dynamic ''states'' characterized either by steady-state solutions for ${\bf a}$ (see Secs.~\ref{iso} and \ref{transition}) and by time-dependent orientational dynamics (Sec.~\ref{nematic}). 
The corresponding total run times
vary between $500$ and $150000$ time steps. 
\section{Dynamics at controlled shear stress}
\label{results}
In the following paragraphs we present numerical results obtained at various dimensionless temperatures $\theta$ around the I-N transition of the equilibrium (unsheared) system.
The coupling parameter $\lambda_{\mathrm{K}}$ appearing in Eqs.~(\ref{eq:T5Dyn})
is set to $1.25$. This choice is motivated by previous investigations at constant shear rate \cite{Rien02,Rien02a} where, at $\lambda_{\mathrm{K}}=1.25$,
a variety of dynamics ''states'' including stationary, oscillatory, and chaotic solutions have been found. Here we include such constant-$\Gamma$ results as a reference. 
For the constant-stress calculations we choose 
$\eta_{\mathrm{iso}}=0.1$ and $\tau_{\mathrm{g}}=1$ [see Eq.~(\ref{dyngam})], if not stated otherwise. Finally, the parameter 
$\kappa$ in Eqs.~(\ref{eq:T5Dyn}) is set to zero. Indeed, earlier studies at constant $\Gamma$ indicate that the actual value of $\kappa$ 
has only minor importance for the dynamic behavior \cite{Hess95}.
\subsection{Isotropic phase}
\label{iso}
\begin{figure}
\includegraphics[width=8cm]{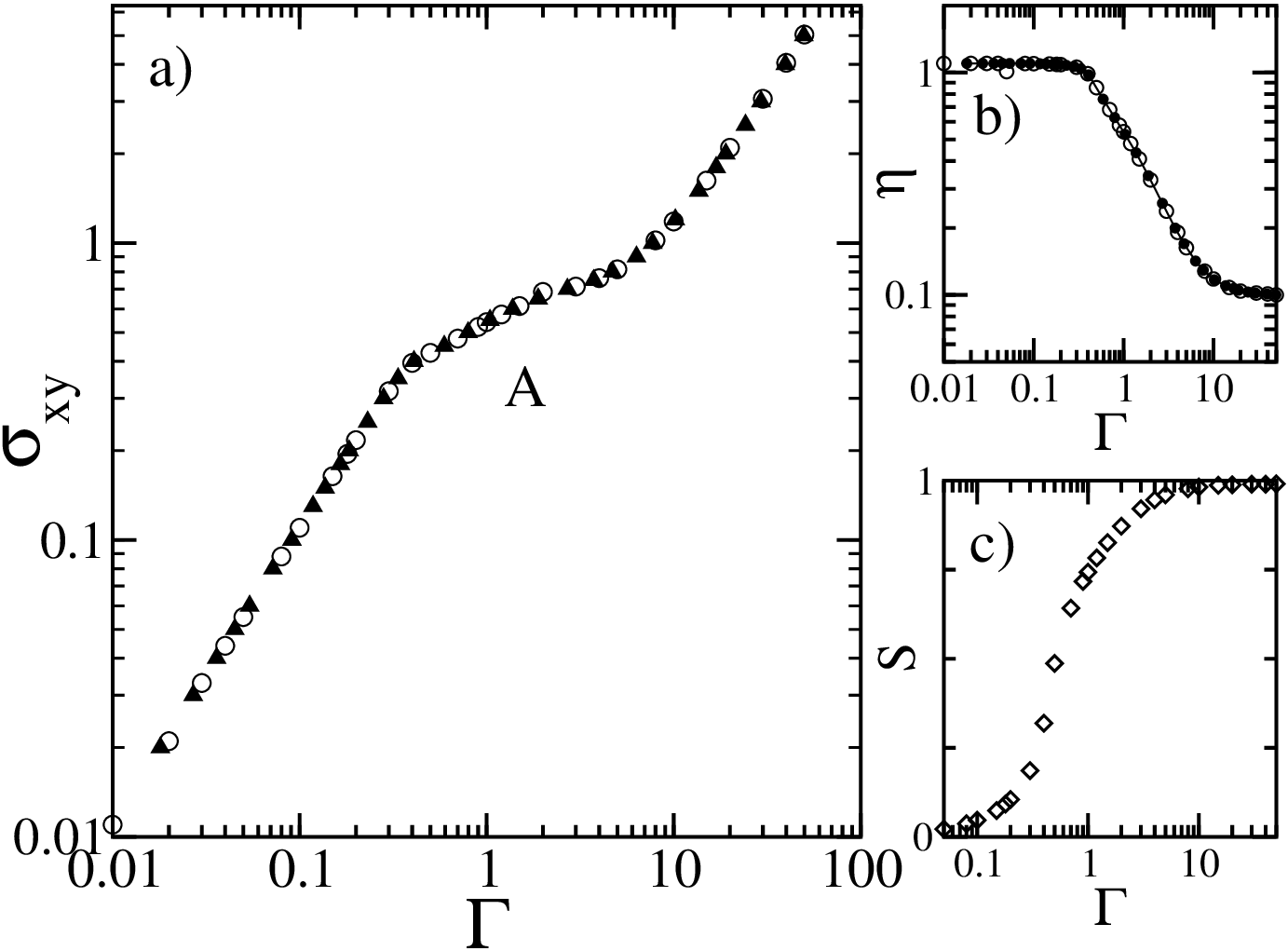}
 \caption{\label{sigma_175}
a) Shear stress $\sigma_{\mathrm{xy}}$ as a function of the shear rate $\Gamma$ (''constitutive curve")
at temperature $\theta = 1.75$, corresponding to the isotropic phase of the equilibrium system.
Results from calculations at fixed $\Gamma$ ($\sigma_{\mathrm{xy}}$) are indicated by open circles (filled triangles).
Parts b) and c) show the corresponding behavior of the viscosity $\eta$ and the Maier-Saupe order parameter $S$, respectively. }
\end{figure}

We start by considering the temperature $\theta=1.75$ which is well within the isotropic phase of the unsheared system (i.e., ${\bf a}|_{\Gamma=0}=0$).
Under such conditions, application of an external shear flow
yields a stationary, uniaxial alignment of the liquid-crystal particles in the $x$-$y$ plane (that is, the flow plane) of the system. 
Within our theoretical description, shear-induced alignment (indicated by an "A" in the following)
is reflected by the quantities $a_0$, $a_1$ and $a_2$ approaching non-zero, constant values, whereas $a_3=a_4=0$. The corresponding ''constitutive'' curve 
$\sigma_{\mathrm{xy}}(\Gamma)$ is shown in Fig.~\ref{sigma_175}a), where we have included data sets from both, 
constant strain- and constant-stress calculations.
In both cases, the system quickly reaches (after a few hundred time steps) a unique, stationary state, independent of the initial
conditions. Moreover, the results from the two types of calculations are quantitatively consistent to a very high accuracy. 
Specifically, one observes from Fig.~\ref{sigma_175}a) a
monotonic, continuous, and non-linear (i.e., non-Newtonian) increase of the stress with increasing shear rate 
(note the double-logarithmic scale). The corresponding (dimensionless) viscosity
$\eta=\sigma_{\mathrm{xy}}/\Gamma$ is plotted in Fig.~\ref{sigma_175}b). The decrease of $\eta$ from its equilibrium value ($\eta_{\Gamma=0}$=1.1) to
its high-shear limit ($\eta_{\Gamma\rightarrow\infty}=\eta_{\mathrm{iso}} = 0.1$) reflects a strongly pronounced shear-thinning, which is quite typical for complex fluids
sheared within the isotropic phase \cite{FIEL07}. Finally, the degree of shear-induced (uniaxial) ordering as measured
by the Maier-Saupe parameter $S(\Gamma)$
is plotted in
Fig.~\ref{sigma_175}c). With increasing $S$
the corresponding director aligns more and more along the $y$-direction, that is, the direction determined by the flow velocity. 

\begin{figure}
\includegraphics[width=8cm]{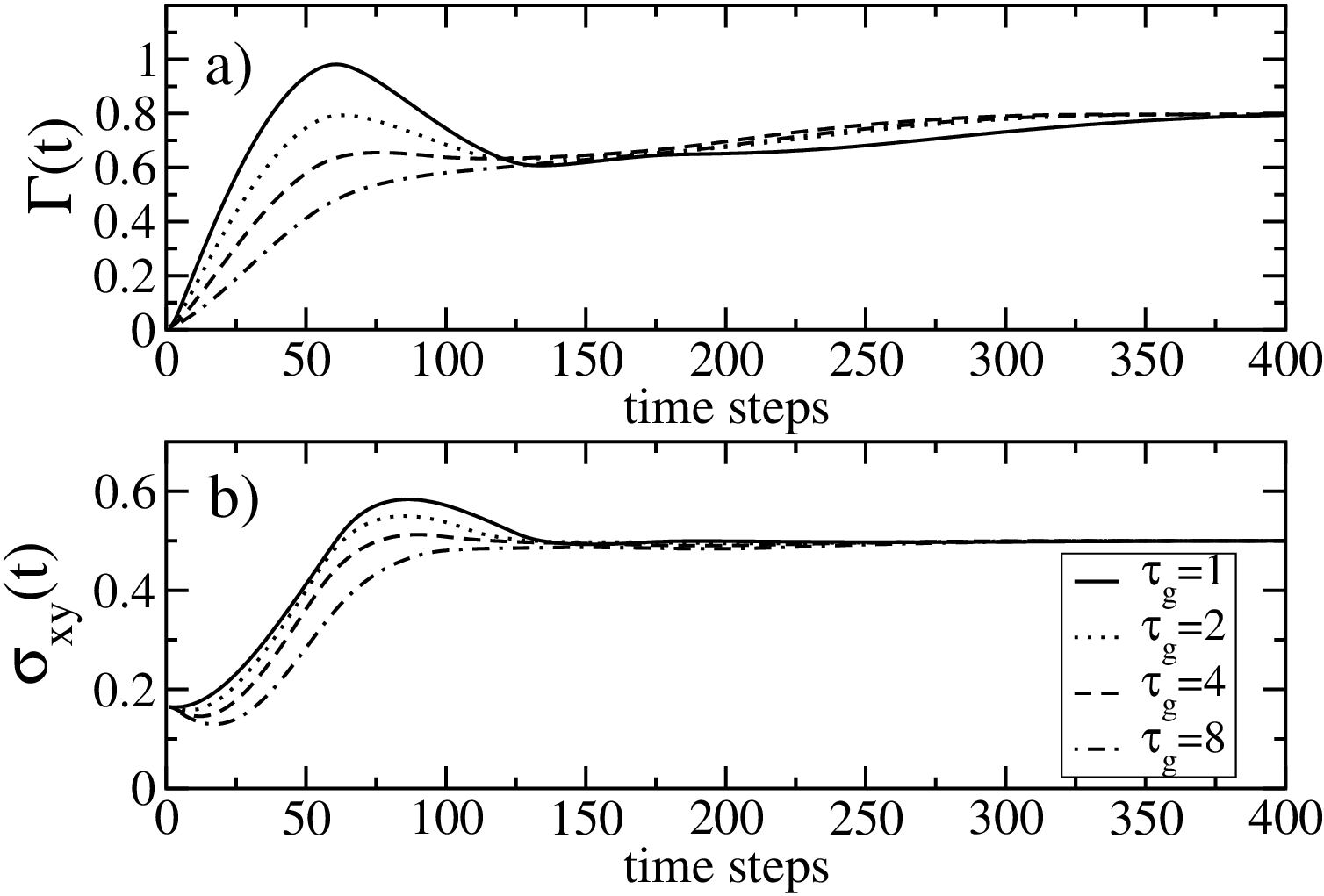}
 \caption{\label{time_175}
 Instantaneous shear rate (a) and shear stress (b) as functions of time under controlled stress conditions ($\sigma_{\mathrm{xy}}^{\mathrm{imp}}=0.5$) at $\theta=1.75$. Included are results
 for four relaxation times $\tau$. The time steps resulting from our adaptive algorithm are $\Delta t\approx 0.3$ for all values of $\tau$ considered.}
\end{figure}

From a more technical point of view it is interesting to note that, at the temperature considered,
the results obtained at fixed $\sigma_{\mathrm{xy}}$
are practically independent  of the choice of the control time chosen in Eq.~(\ref{dyngam}) (as tested for values of $\tau_{\mathrm{g}}$ between $0.1$ and  $100$). 
The precise value of $\tau_{\mathrm{g}}$ merely affects the amount of time required to reach the stationary limit.  This is illustrated in Fig.~\ref{time_175}a),
where we plot the time-dependence of the instantaneous shear rate $\Gamma(t)$ at fixed, imposed
stress $\sigma_{\mathrm{xy}}^{\mathrm{imp}}=0.5$ for various delay times $\tau_{\mathrm{g}}$.
The larger $\tau_{\mathrm{g}}$, the slower is the approach of $\Gamma(t)$ towards its asymptotic value ($\langle\Gamma\rangle\approx 0.79$), as one might have
expected from the structure of the control equation~(\ref{dyngam}).
The corresponding time dependence of the instantaneous shear stress $\sigma_{\mathrm{xy}}(t)$ is plotted in Fig.~\ref{time_175}b). It is seen that 
$\sigma_{\mathrm{xy}}(t)$ relaxes somewhat faster as compared to $\Gamma(t)$, independent of the actual value of $\tau_{\mathrm{g}}$.
\subsection{Shear-induced isotropic-nematic transition} 
\label{transition}
\begin{figure}
\includegraphics[width=8cm]{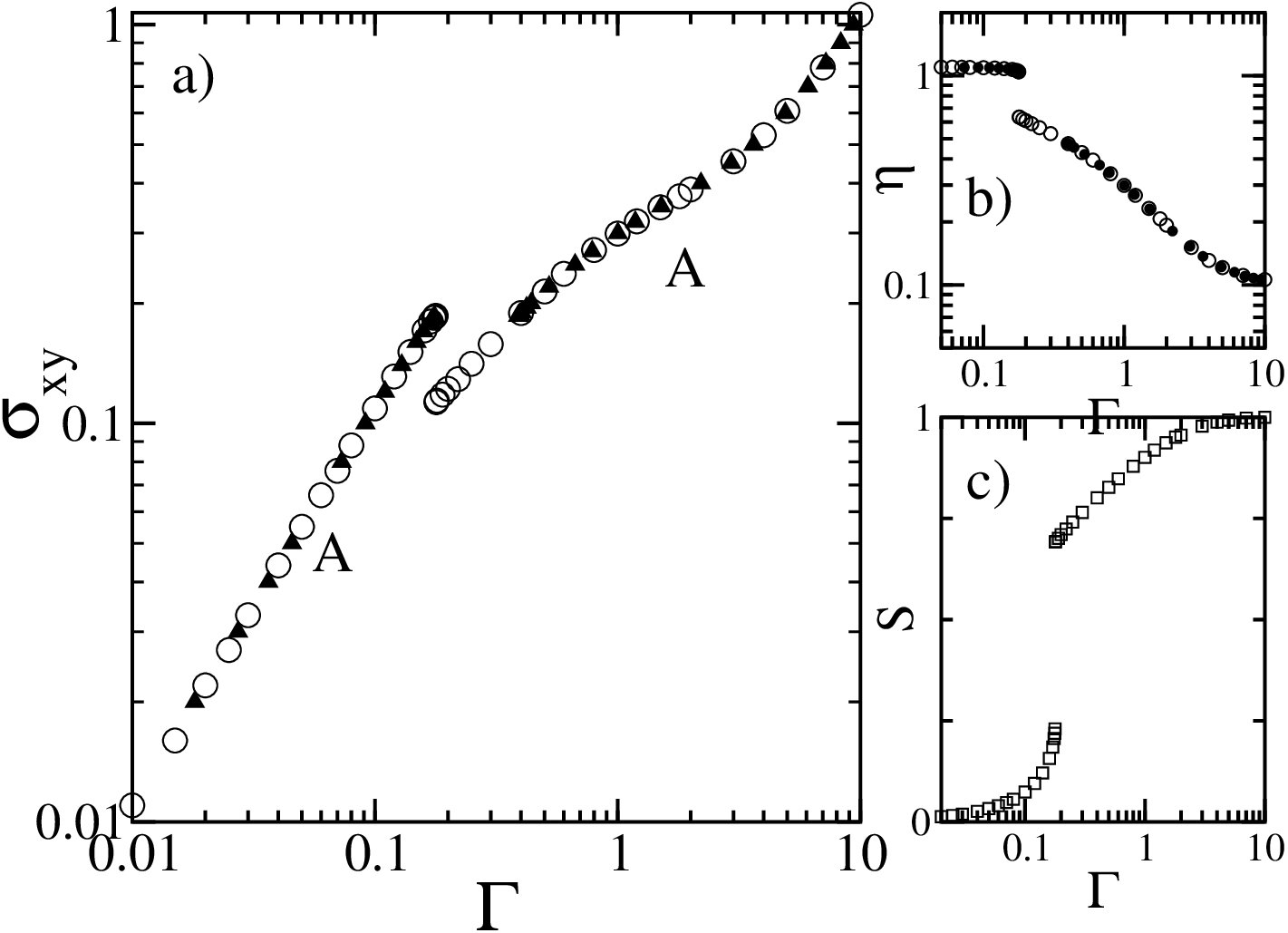}
 \caption{\label{sigma_125}
 Same as Fig.~\ref{sigma_175}, but at temperature $\theta = 1.25$ close above the I-N transition of the equilibrium system. }
\end{figure}

We next consider the temperature $\theta=1.25$. The corresponding equilibrium system is still globally isotropic, but much closer to the first-order I-N transition (occurring at $\theta=1$) than the state considered in Sec.~\ref{iso}.  As a consequence of the nearby equilibrium transition, 
shearing induces a {\it non-equilibrium} transition from a ''paranematic'' state characterized by weak orientational ordering 
into a state with pronounced stationary and (almost) uniaxial nematic order. We note that the
equilibrium system has a metastable nematic phase only for lower temperatures, that is, for $\theta<9/8=1.125$). The shear-induced transition is illustrated in Figs.~\ref{sigma_125}a)-c), where we plot 
the stress-strain relation at $\theta=1.25$ together with corresponding results for the viscosity and the Maier-Saupe parameter.
For imposed shear rate, the non-equilibrium transition occurs $\Gamma\approx 0.178$ where we observe pronounced discontinuities 
in all of the three quantities investigated. Specifically, the jump-like increase of $S$ is accompanied by jumps in $\sigma_{\mathrm{xy}}$ (and, consequently, $\eta$) towards smaller values, indicating a sudden shear-thinning 
of the system. We note that the occurrence of the shear-induced transition has been
realized, on a theoretical basis, already quite some time ago \cite{Hess76a,Olmsted92}. A recent experimental realization involves colloidal suspensions of tobacco viruses \cite{Lettinga05}.

Coming back to our model system, the data in Fig.~\ref{sigma_125}a) imply that there are values of $\sigma_{\mathrm{xy}}$ where two dynamic states characterized by different shear rates
$\Gamma$ can ''coexist''.
Interestingly, when controlling the shear stress [see black triangles in Fig.~\ref{sigma_125}a)], the system remains in the branch with lower $\Gamma$ up to largest value of $\sigma_{\mathrm{xy}}$ where coexistence is possible. This observations turns out to be essentially independent of the details of the stress control, i.e. the control time
$\tau_{\mathrm{g}}$ (and the initial value for $\Gamma$). 
This rather robust behavior is in sharp contrast to the non-equilibrium transitions at lower temperatures to be discussed next. 
\subsection{Shearing within the nematic phase}
\label{nematic}
\begin{figure}
\includegraphics[width=8cm]{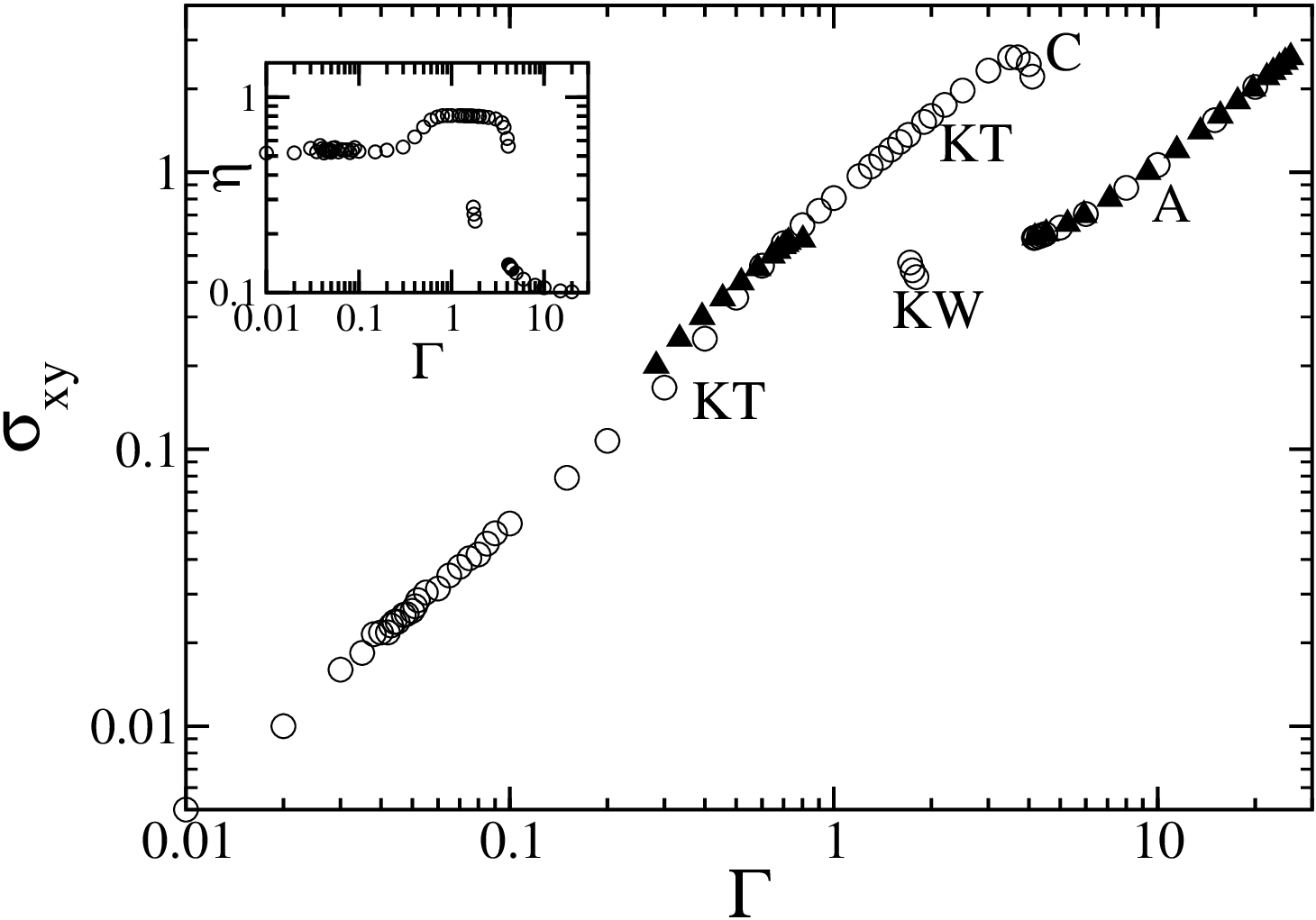}
 \caption{\label{sigma_0}
Shear stress $\sigma_{\mathrm{xy}}$ as a function of the shear rate $\Gamma$ at temperature $\theta = 0$, within the nematic phase of the equilibrium system.
Results from calculations at fixed $\Gamma$ ($\sigma_{\mathrm{xy}}$) are indicated by open (filled) circles.
Calculations at fixed stress have been performed at $\tau_{\mathrm{g}}=1$.
The abbreviations ''KT'', "W", "C", and "A" indicate the orientational behavior and are explained in the main text.
The inset shows the corresponding behavior of the viscosity. }
\end{figure}

\begin{figure}
\includegraphics[width=8cm]{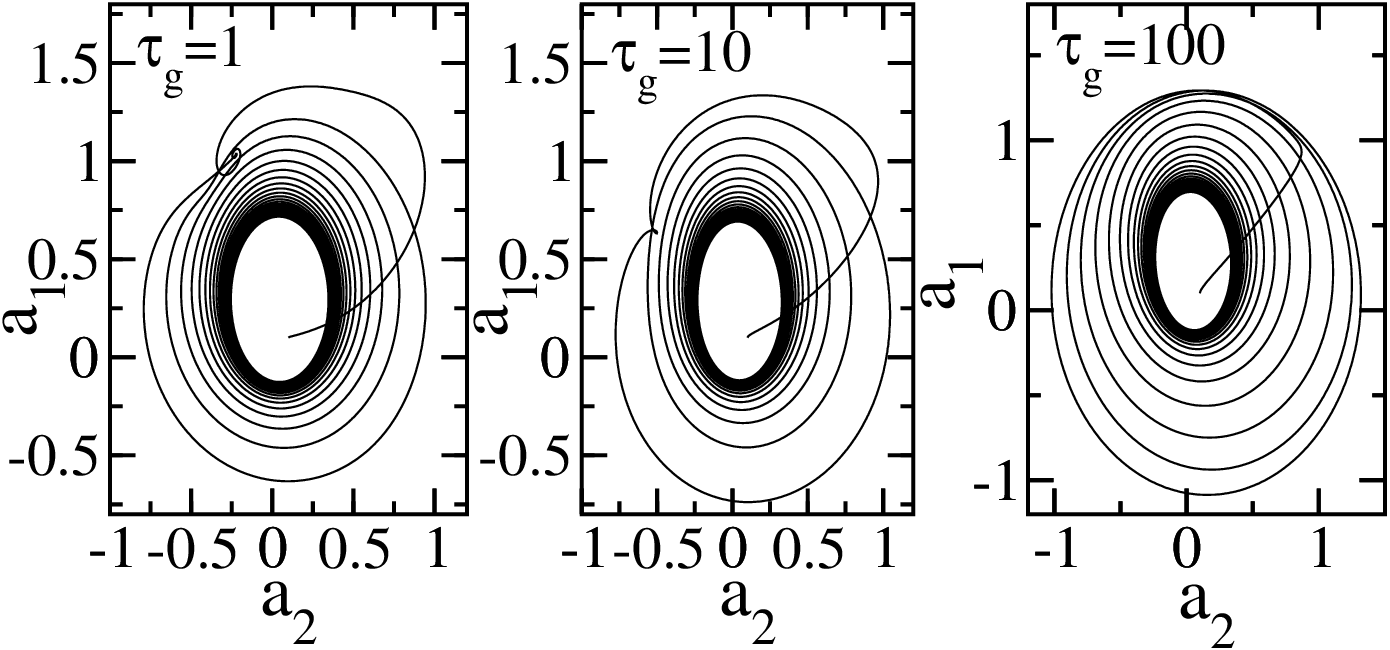}
 \caption{\label{a12_0}
 Phase portraits of the in-plane components of the pressure tensor, $a_2$ versus $a_1$, for different
 control times $\tau_{\mathrm{g}}$. The average shear stress is fixed at $\sigma_{\mathrm{xy}}^{\mathrm{imp}}=0.575$.}
 \end{figure}

\begin{figure}
\includegraphics[width=8cm]{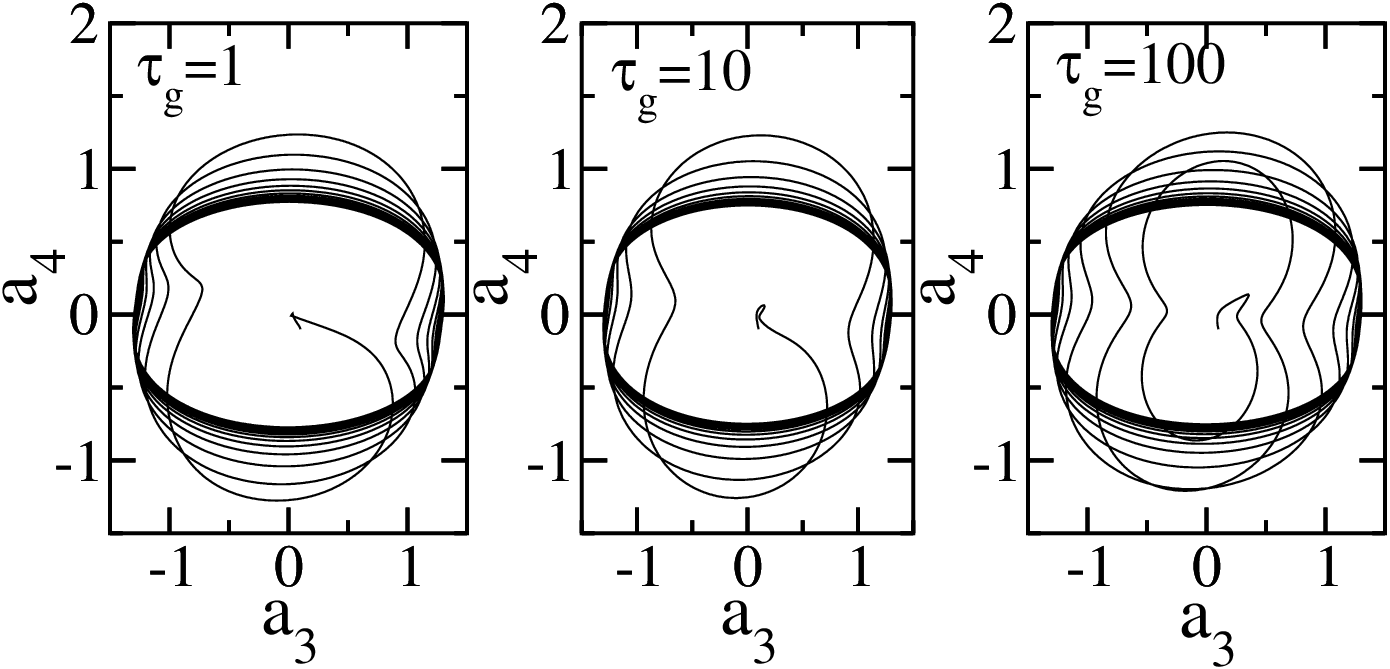}
 \caption{\label{a34_0}
 Same as Fig.~\ref{a12_0}, but for the out-of plane components $a_4$ versus $a_3$.}
 \end{figure}

As a representative example for the system's low-temperature dynamics we consider the 
temperature $\theta = 0$. Here, the nematic state is globally stable already at zero shear, and we are at the lower limit of (meta-)stability
of the isotropic phase. Our results for the shear stress as function of $\Gamma$ are displayed in Fig.~\ref{sigma_0}.
All calculations have been started from almost random initial states (see Sec.~\ref{numerics}).
Earlier numerical studies \cite{Rien02} with controlled shear rate and $\lambda_{\mathrm{K}}=1.25$ have already revealed the occurrence of various types of time-dependent dynamics
of the nematic director. These include "symmetry-breaking'' states where the director rotates out of the shear plane
(i.e., $a_0(t)>0$, $a_3(t)\neq 0$, $a_4(t)\neq 0$). Our present results at constant $\Gamma$ are fully consistent with those in \cite{Rien02}.
Specifically, we observe for low shear rates ($\Gamma\lesssim 1.7$) so-called kayaking-tumbling (KT) motion, where director performs out-of-plane oscillations, and its projection 
onto the shear plane describes an ellipse (i.e., tumbling). Corresponding orbits illustrating the time dependence of $a_1$-$a_4$ are shown
in Figs.~\ref{a12_0} and \ref{a34_0}. Increasing $\Gamma$ there exists a (small) window
($1.7\lesssim \Gamma \lesssim 1.9$) characterized by kayaking-wagging (KW) motion, 
where the director performs out-of-plane oscillations while its projection in the shear plane describes wagging. 
From Fig.~\ref{sigma_0} we see that the transition into the KW state is supplemented by a sudden decrease of the shear stress. Upon further
increase of $\Gamma$ we are re-enterting the KT state until $\Gamma\approx 3.5$. In the subsequent range of shear rates the system has chaotic (C) features, with
the long-time dynamics being strongly dependent on the initial conditions and the largest Lyaponov exponent being positive (for detailed discussion of the
chaotic regime, see Refs.~\cite{Rien02,Rien02a}). Finally, for shear rates $\Gamma\gtrsim 4.15$ there is a further transition
into a flow-aligned (A) state where the director is arrested to a constant direction within the shear plane.

\begin{figure}
\includegraphics[width=8cm]{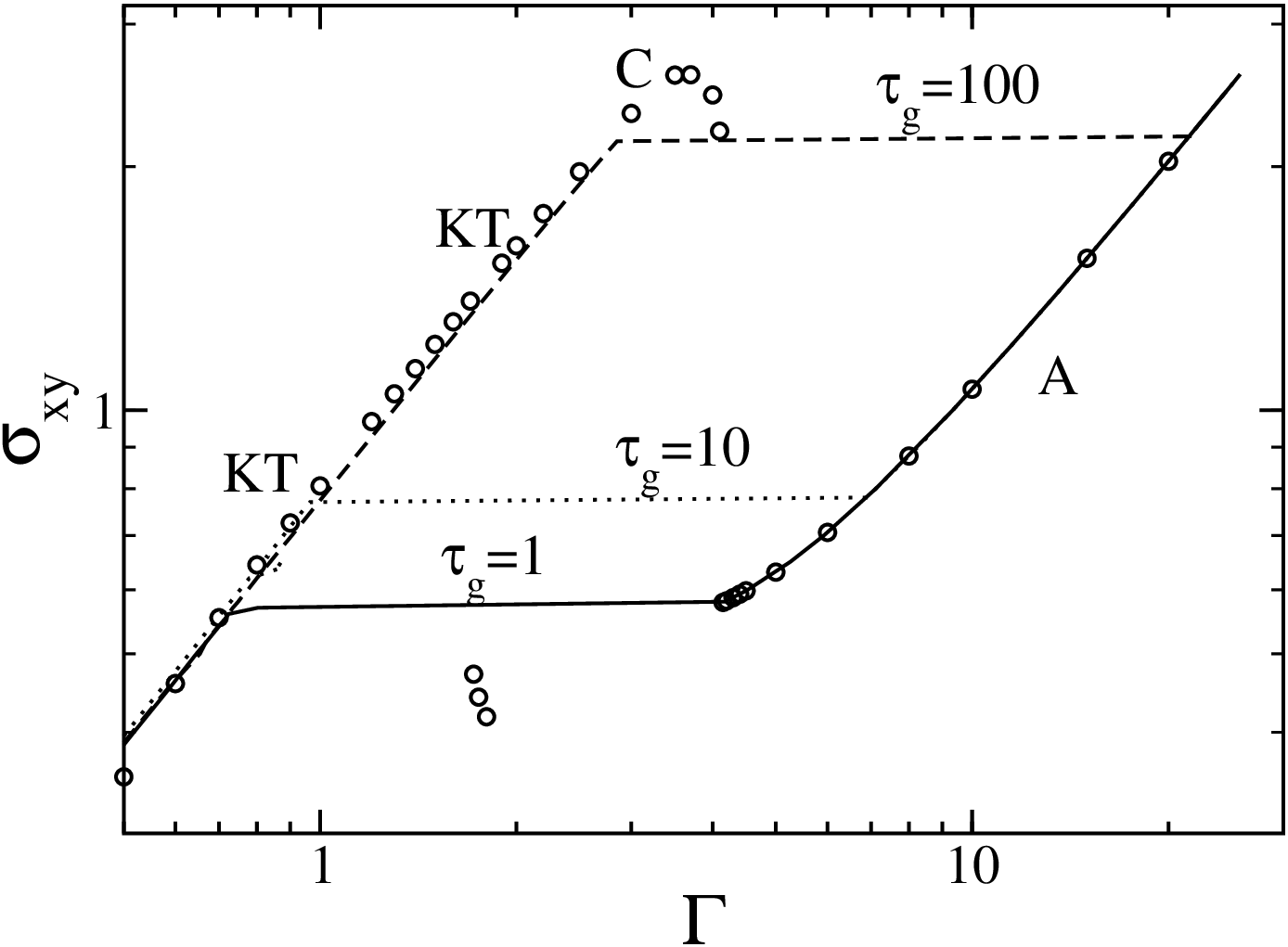}
 \caption{\label{sigma_tau_0}
 Enlarged version of the stress-strain curve in Fig.~\ref{sigma_0}. The lines indicate the result of constant-stress calculations 
 with different values of the relaxation time $\tau_{\mathrm{g}}$.}
\end{figure}

Remarkably, not all of these states are observable when we take the shear stress as a control variable. This can be seen from the data points indicated by (black) triangles
in Fig.~\ref{sigma_0}. Specifically, our controlled-stress calculations do neither yield the wagging motion, nor the complex, chaotic behavior. Instead, the system jumps
directly from the KT into the shear-aligned (A) state. Moreover, it turns out that the {\it position} of this jump in the $\Gamma$-$\sigma_{\mathrm{xy}}$ plane can be manipulated, to some extent,
by the control time $\tau_{\mathrm{g}}$ (indeed, the initial value of the shear rate has much less effect). The role 
of $\tau_{\mathrm{g}}$ is illustrated in Fig.~\ref{sigma_tau_0}, where we present results for various values of $\tau_{\mathrm{g}}$ on an enlarged scale.
The larger the control time, the longer the system remains on the low-$\Gamma$ branch. In this sense, an increase of $\tau_{\mathrm{g}}$ stabilizes the periodic KT-state.
Interestingly, the {\it chaotic} region (C) does not appear in stress-controlled calculations even for the
largest values of $\tau_{\mathrm{g}}$ considered. Indeed, as we will demonstrate in Sec.~\ref{stabilization}, a sudden onset of shear-stress control 
can actually stabilize these chaotic states. 

\begin{figure}
\includegraphics[width=8cm]{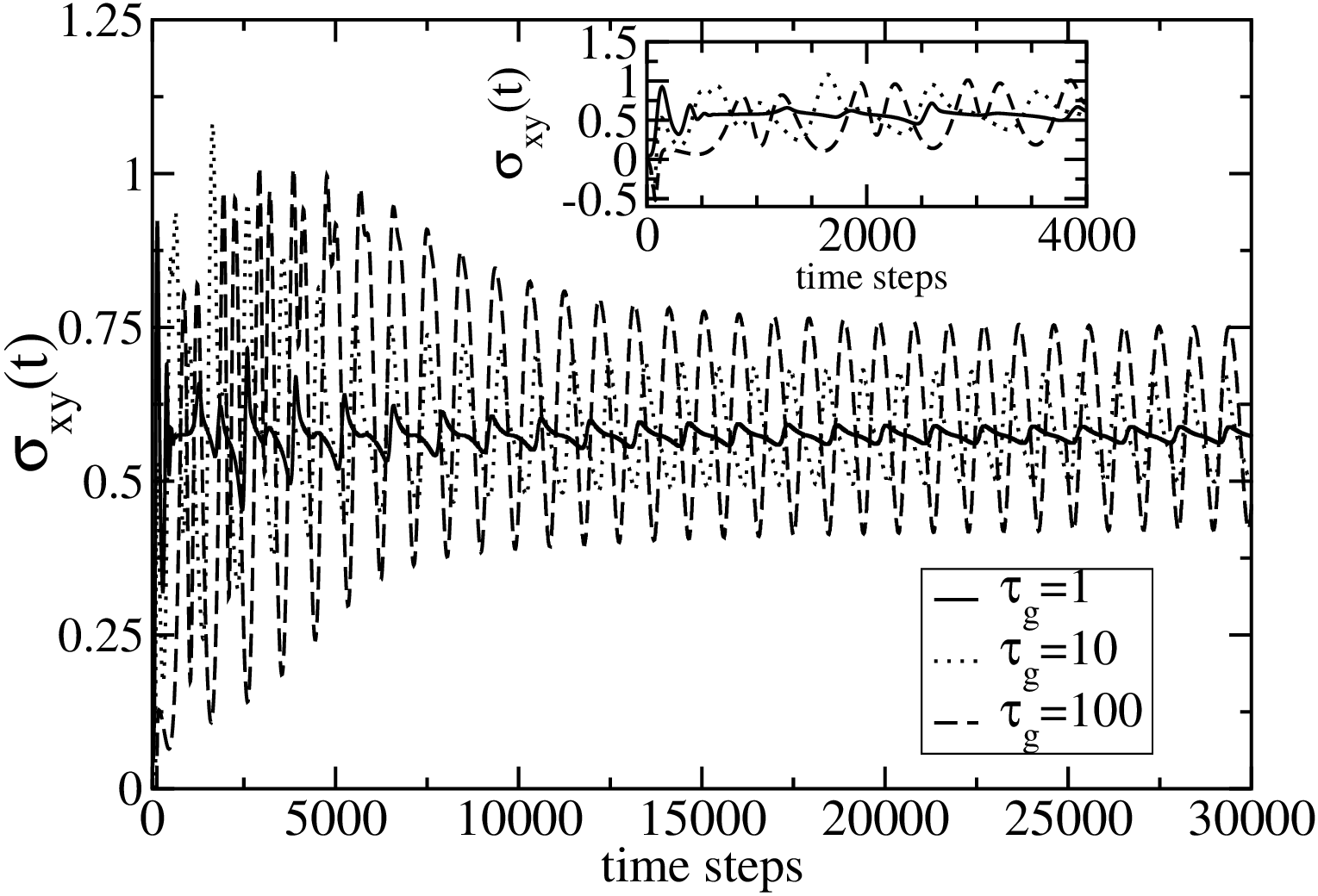}
 \caption{\label{stress_time_0}
Instantaneous stress as function of time for different
 control times $\tau_{\mathrm{g}}$. The average shear stress is fixed at $\sigma_{\mathrm{xy}}^{\mathrm{imp}}=0.575$.
 The resulting time steps are $\Delta t\approx 0.01$ for all values of $\tau$ considered.}
 \end{figure}

\begin{figure}
\includegraphics[width=8cm]{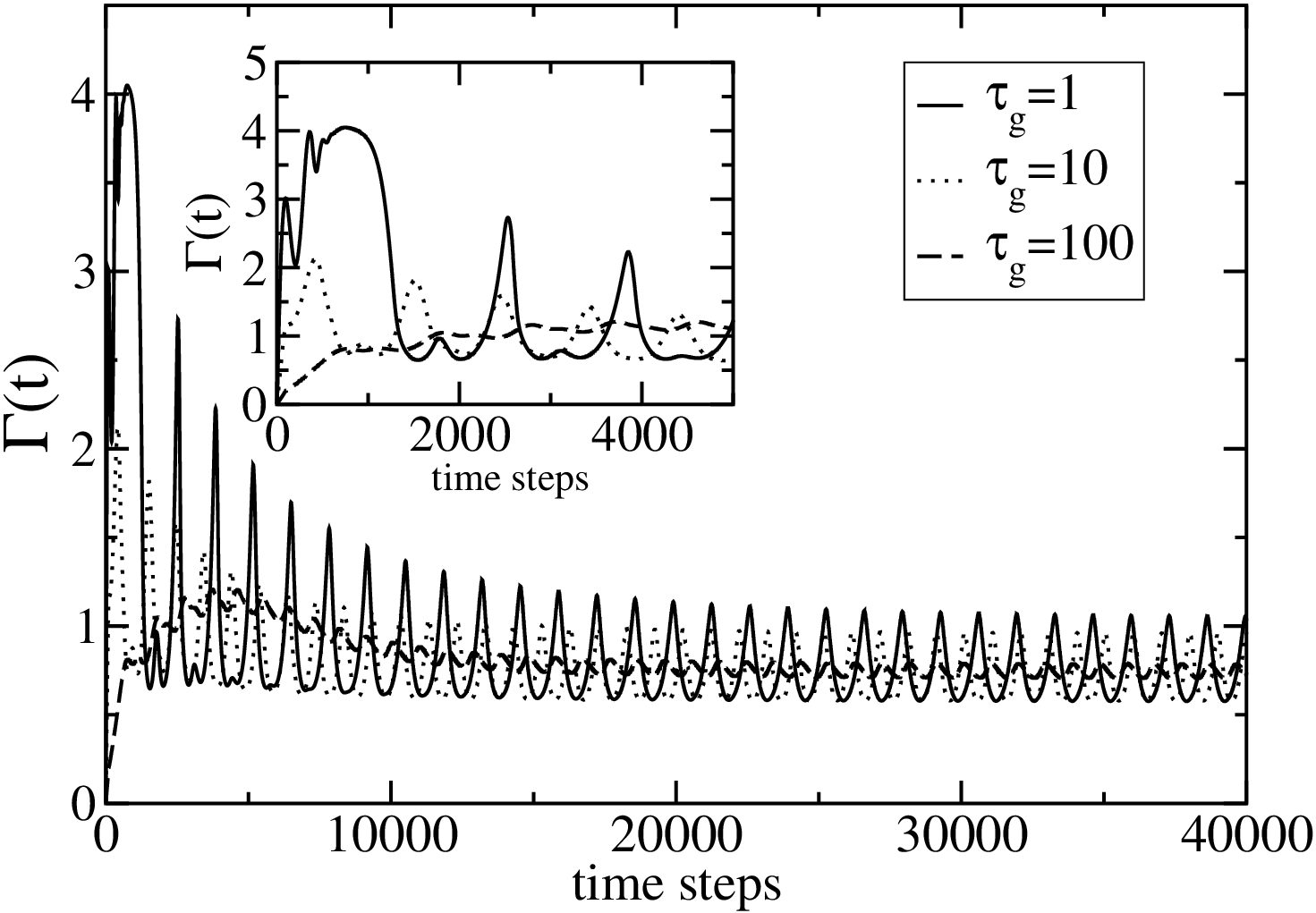}
 \caption{\label{shear_time_0}
 Instantaneous shear rate as function of time for different
 control times $\tau_{\mathrm{g}}$. Parameters as in Fig.~\ref{stress_time_0}.}
 \end{figure}

Even away from the KT-A transition the dynamics in the stress-controlled, nematic system depends significantly on the control time, $\tau_{\mathrm{g}}$. This is illustrated
in the subsequent figures~\ref{stress_time_0}-\ref{a34_0}, where we consider various time-dependent quantities
at an imposed stress of  $\sigma_{\mathrm{xy}}^{\mathrm{imp}}=0.575$. According to Fig.~\ref{sigma_0}, this value corresponds to a state within the KT regime (and below
the KT-A transition). To start with, the time-dependence of the {\it instantaneous} stress $\sigma_{\mathrm{xy}}(t)$ and that of the shear rate $\Gamma(t)$ are displayed
in Figs.~\ref{stress_time_0} and \ref{shear_time_0}, respectively.
For all control times considered, both quantities display pronounced oscillations even in the long-time limit. In particular, $\sigma_{\mathrm{xy}}(t)$ reaches
its externally imposed value of
$\sigma_{\mathrm{xy}}^{\mathrm{imp}}=0.575$ only on the {\it average} [see Fig.~\ref{stress_time_0}]. The amplitude of the oscillations strongly depends on $\tau_{\mathrm{g}}$. Specifically, an increase of the
control times tends to damp out the oscillations. Similar behavior is seen in Fig.~\ref{shear_time_0} for the shear rate $\Gamma(t)$.

The orientational dynamics corresponding to the KT state at $\sigma_{\mathrm{xy}}^{\mathrm{imp}}=0.575$ is illustrated in
Figs.~\ref{a12_0} and \ref{a34_0}, where we show phase portraits (orbits) of both, the in-plane motion ($a_2$ as function of $a_1$)
and the out-of-plane motion ($a_4$ as function of $a_3$).
The ellipse-shaped limit cycle characterizing the functions $a_2(a_1)$ [see Fig.~\ref{a12_0}] 
is typical for the KT motion, where the nematic director displays tumbling motion in the shear plane. A closer analysis
of the time-dependence of the individual components reveals that the {\it transient} behavior observed in the period between $\approx 400$ and $\approx 1400$ time steps (i.e.,
well before the asymptotic KT state is reached) somewhat depends
on the control time, $\tau_{\mathrm{g}}$. Specifically, at $\tau_{\mathrm{g}}=1$ there is transient flow alignment where $a_0$, $a_1$, $a_2$ are constant, and 
$a_3\approx 0$, $a_4\approx 0$. At $\tau_{\mathrm{g}}=10$ one finds instead
a short period of in-plane tumbling, where $a_1$ and $a_2$ become time-dependent. These differences in the alignment dynamics are mirrored by corresponding differences in the 
transient behavior of the instantaneous shear rate and stress, respectively
(see insets in Figs.~\ref{stress_time_0} and \ref{shear_time_0}). At larger times the orbits related to different values of the control time $\tau_{\mathrm{g}}$ look rather similar. 
This is in accordance to earlier findings on the robustness of the orientational
dynamics with respect to time-dependent perturbations of the shear rate \cite{Heidenreich06}.
\subsection{Stabilization of chaotic states}
\label{stabilization}
\begin{figure}
\includegraphics[width=8cm]{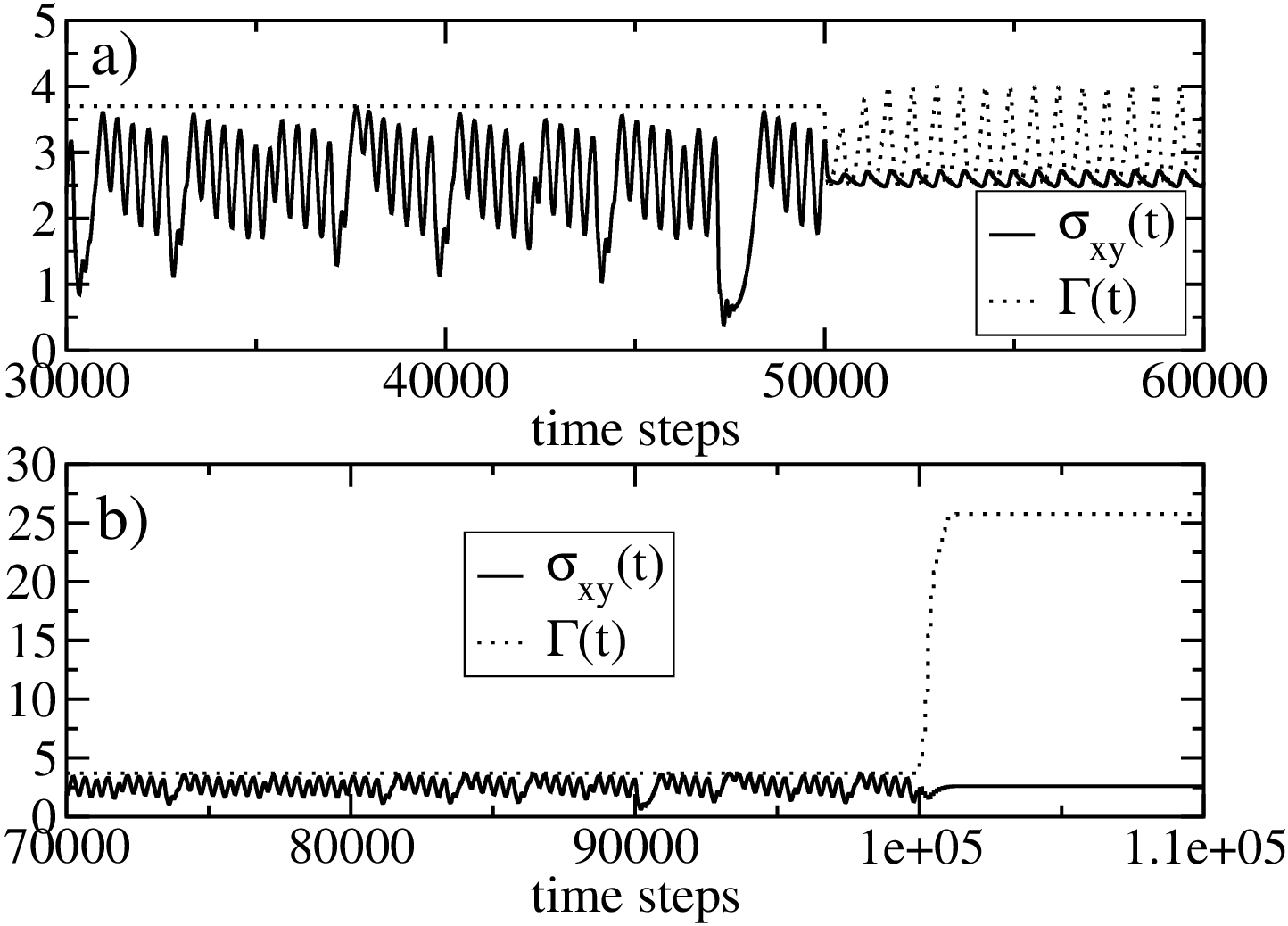}
 \caption{\label{stress_switch}
 Shear stress and shear rate as functions of time in systems where the control method switches
 from fixed shear rate ($\Gamma=3.7$) to fixed shear stress ($\sigma_{\mathrm{xy}}^{\mathrm{imp}}=2.6$, $\tau_{\mathrm{g}}=1$).
 The switch occurs at
 a) $N_{\mathrm{s}}=5\times 10^4$ time steps, b) $N_{\mathrm{s}}=1\times 10^5$ time steps. 
 The resulting time steps are $\Delta t\approx 0.01$ for both values of $N_{\mathrm{s}}$.}
 \end{figure}

\begin{figure}
\includegraphics[width=8cm]{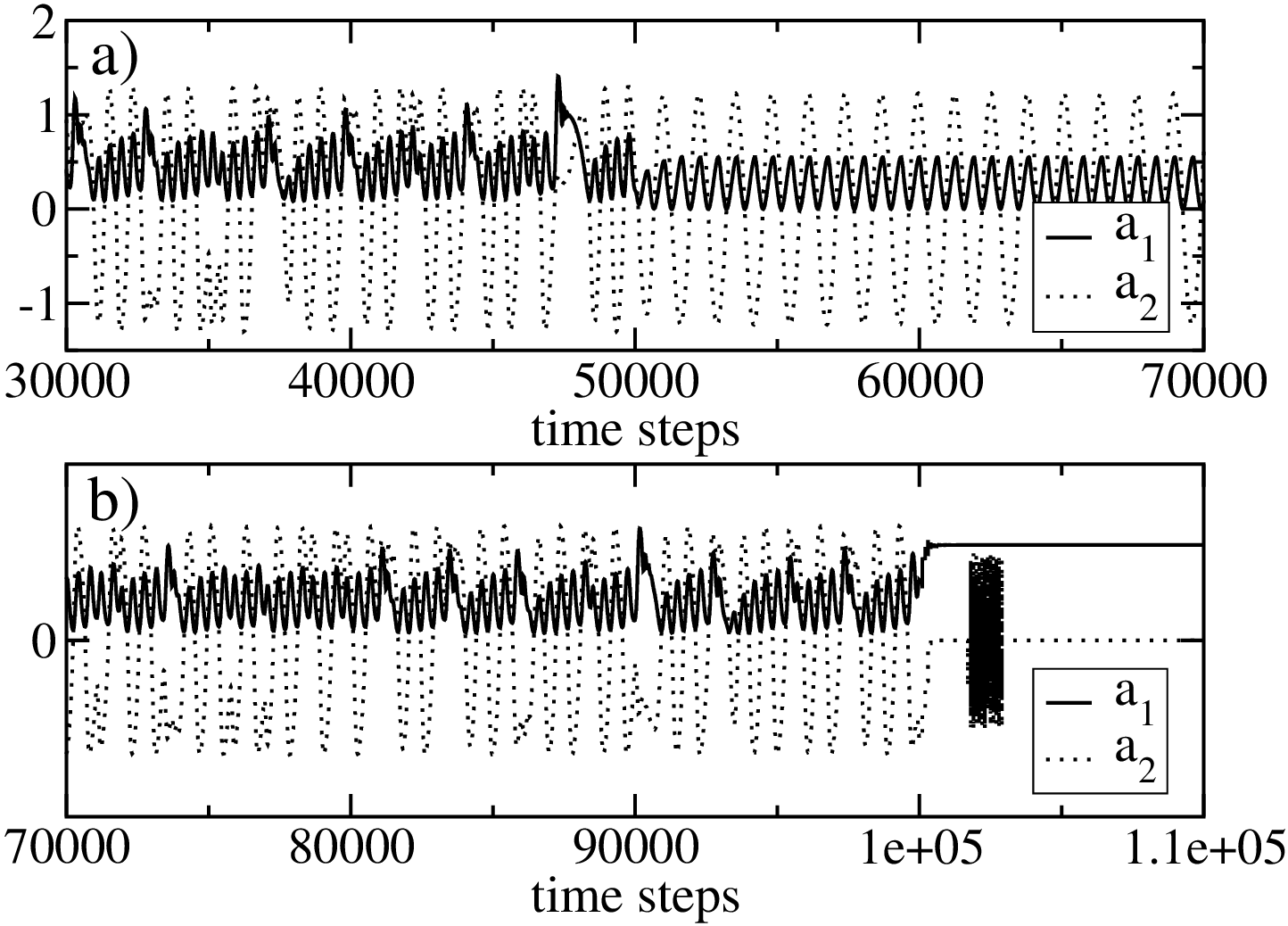}
 \caption{\label{a1a2_switch}
 Same as Fig.~\ref{stress_switch}, but for the components $a_1$ and $a_2$ of the alignment tensor.}
 \end{figure}

 An interesting feature of our results within the nematic phase is that under controlled stress conditions, the transition from the KT into the shear-aligned state occurs
 before the system enters the chaotic (C) regime observed in calculations at fixed $\Gamma$ (see Fig.~\ref{sigma_0}). 
 This suggests that stress control might have a {\it stabilizing} influence on
 shear-rate-induced chaotic states. To explore this conjecture we have performed test ''experiments'' where the control method was {\it switched} from fixed shear rate to fixed
 stress after a predefined number $N_{\mathrm{s}}$ of time
 steps. That is, for $N<N_{\mathrm{s}}$ the orientational dynamics was calculated via the five equations~(\ref{eq:T5Dyn}),
 whereas at $N>N_{\mathrm{s}}$ these equations were used
 together with Eq.~(\ref{dyngam}). As an representative example, we consider the behavior obtained at $\Gamma=3.7$ ($N<N_{\mathrm{s}}$) and
 $\sigma_{\mathrm{xy}}^{\mathrm{imp}}=2.6$ ($N>N_{\mathrm{s}}$). Numerical results for the instantaneous shear stress and -rate,
 are displayed in Fig.~\ref{stress_switch}, where we have included data for two values of $N_{\mathrm{s}}$.
The corresponding behavior of the in-plane components of the director, $a_1$ and $a_2$, is shown in Fig.~\ref{a1a2_switch}.
In both figures, one clearly recognizes a drastic change of the dynamics upon the switch from shear- to stress control. Before the switch one observes
highly irregular dynamics of all non-fixed quantities, that is $\sigma_{\mathrm{xy}}(t)$, $a_1(t)$, $a_2(t)$ (and the same holds for the other components of
the tensor ${\bf a}$). Closer results reveals that the system at fixed $\Gamma=3.7$ is indeed in a chaotic state \cite{Rien02,Rien02a}. After the switch, that is, under controlled stress, the dynamics becomes regular. Interestingly, the {\it character} of this final state depends on both, the starting point ($N_{\mathrm{s}}$)
of the stress control, and the corresponding delay time $\tau_{\mathrm{g}}$.
For the choices $N_{\mathrm{s}}=5\times 10^4$ and $\tau_{\mathrm{g}}=1$ [see Figs.~\ref{stress_switch}a) and \ref{a1a2_switch}a)], 
the systems displays KT dynamics characterized by oscillatory (yet regular) motion of all components
$a_1$-$a_4$ and corresponding oscillations of $\Gamma(t)$. The average shear rate is $\langle\Gamma\rangle\approx 3.1$, which is again a typical value for the KT regime as seen
in Fig.~\ref{sigma_0}. Performing then the switch at the later time $N_{\mathrm{s}}=1\times 10^5$ 
(and $\tau_{\mathrm{g}}=1$) [see Figs~\ref{stress_switch}b) and \ref{a1a2_switch}b)], all of the displayed
quantities become {\it constant}. This signals that we are entering a shear aligned (A) state, characterized by a shear rate of $\langle\Gamma\rangle\approx 25.8$. It is noteworthy that the same
A state is achievable with the choices $N_{\mathrm{s}}=5\times 10^4$ and $\tau_{\mathrm{g}}=10$ (or $100$).
Taken altogether, these results reveal that stress
control can indeed stabilize the dynamics in systems characterized by unstable motion without control. Moreover, the character of the final state is tunable by the details of the control scheme,
that is in our case, the starting time of the stress control, and the corresponding delay time.
\section{Conclusions}
\label{conclude}
In this article we have proposed a mesoscopic theory to investigate the orientational dynamics and rheology of sheared complex fluids
at fixed, externally imposed shear stress $\sigma_{\mathrm{xy}}$. 
Our strategy to control $\sigma_{\mathrm{xy}}$ consists  of supplementing
the previously established equations of motion for the orientational order parameters \cite{Rien02a}
by a relaxation equation 
for the shear rate, $\Gamma$. The speed of this relaxation can be tuned by a control ("delay") time. Based on the resulting six-dimensional dynamical system 
we have numerically investigated sheared fluids of rod-like particles at several temperatures
around the I-N transition at $\Gamma=0$.  
At high temperatures, where the equilibrium system is isotropic, shear stress control yields the same asymptotic behavior (that is, flow alignment) as that encountered at fixed $\Gamma$. Within the low-temperature nematic phase, however, the control variable (stress or strain) influences both, the location
and the {\it type} of bifurcations between different dynamic states. 
This finding is relevant not only from a theoretical point of view but also for experiments; indeed, in rheological set-ups it is often the shear stress rather than the shear rate which 
can be fixed externally. In our model, the stress-controlled nematic fluid is characterized by only one non-equilibrium transition
from a periodic (kayaking-tumbling) into a flow-aligned state, contrary to its $\Gamma$-controlled counterpart which displays, in addition,
wagging and chaotic motion. Furthermore, the dynamics at fixed stress is strongly influenced by the delay time
$\tau_{\mathrm{g}}$. Large values of the latter tend to damp oscillatory motion. Moreover, the delay time
can be used as a parameter to tune the position of the system's transition from a low-$\Gamma$ to a high-$\Gamma$ branch. Finally, we have demonstrated that our control scheme is capable of stabilizing regular (periodic or stationary) states out of chaotic orbits appearing at fixed $\Gamma$.

In the present paper we have restricted ourselves to the investigation of spatially homogeneous states. 
This is clearly an oversimplification in view of experiments pointing to shear-induced inhomogeneities (see, e.g., \cite{LER08,Rendon07,Park09}). However,
as mentioned before, our approach is in principle fully capable of treating inhomogeneities as demonstrated, e.g., in Refs.~\cite{Heidenreich09}
and \cite{God10}, the latter study dealing with a full three-dimensional spatially resolved flow problem. 
Interestingly, there is a recent study of Das {\it et al.} \cite{Das05} suggesting that 
the chaotic orientational behavior of our present system at constant $\Gamma$ transforms into a
spatiotemporal chaos, when spatial fluctuations are allowed in the theory (via gradient terms in the free energy). This suggests that the stress-controlled dynamics reported in present work may also be spatially extended. 
We note, however, that a full description of the spatially resolved problem would include {\it feedback} effects of the orientational motion on the flow \cite{Heidenreich09} (which are neglected
in \cite{Das05}, where $\Gamma$ was assumed to be a constant). Given the complexity of the resulting numerical study we consider the present findings as 
first indicators for the true dynamics and as useful starting points for a more detailed investigation. A further interesting direction would be an extension of the present study
towards polar systems \cite{Heidenreich09,Grandner07}.

Finally, the findings in the present study, particularly those regarding chaotic states, are also interesting in the broader context of control in non-linear dynamic systems \cite{Schoellbuch}. In particular, there 
is currently an intense research on systems such as lasers \cite{DAH08b}, neural systems \cite{DAH08}, excitable media \cite{Schlesner08,Dah08c},
and Brownian ratchet systems \cite{Craig08,Hennig09}, where unstable dynamic behavior is controlled
via {\it time-delayed feedback control} (TDAS) \cite{Pyragas92}. In this framework the equations of motion are supplemented by ''control force" terms 
involving differences of an appropriate system variable (or output quantity) at a time $t$ and an earlier time $t-\tau$. The control force thus vanishes
when a specified stationary or periodic state is reached. This strategy is somewhat different from the present approach where the control is realized by an {\it additional} equation
of motion for the shear rate $\Gamma$, which is varied until $\sigma_{\mathrm{xy}}$ approaches an externally imposed value. Despite these differences, a common feature of our approach
and the TDAS method is that the time parameter ($\tau$ or $\tau_{\mathrm{g}}$, respectively) plays a crucial role for the asymptotic dynamics
(in fact, the present scheme could be viewed as feedback control with additional low-pass filtering involving an exponentially distributed delay
\cite{HOV05}). Indeed, in our model the actual value of $\tau_{\mathrm{g}}$ decides which type of non-chaotic state the unstable system selects when starting the control within the chaotic regime. 
These similarities suggest that control concepts established in non-linear systems, such as TDAS, may be very fruitful not only for a deeper understanding of the rheological properties, but also for the orientational dynamics.
Work in this direction is under way.

\section*{Acknowledgement}
We thank Eckehard Sch\"oll for stimulating discussions.

\end{document}